\documentclass[aps,prl,showpacs,preprintnumbers,nofootinbib,float,longbibliography,superscriptaddress,notitlepage,twocolumn,tikz]{revtex4-1}

\usepackage{physics}
\usepackage{tikz}
\usepackage{tikz-3dplot}
\usepackage[outline]{contour} % glow around
\colorlet{veccol}{green!50!black}
\colorlet{projcol}{blue!70!black}
\colorlet{myblue}{blue!80!black}
\colorlet{myred}{red!90!black}
\colorlet{mydarkblue}{blue!50!black}

\definecolor{mygreen}{RGB}{0,117,0}
\tikzset{>=latex} % for LaTeX arrow head
\tikzstyle{proj}=[projcol!80,line width=0.08] %very thin
\tikzstyle{area}=[draw=veccol,fill=veccol!80,fill opacity=0.6]
\tikzstyle{vector}=[-stealth,myblue,thick,line cap=round]
\tikzstyle{unit vector}=[->,veccol,thick,line cap=round]
\tikzstyle{dark unit vector}=[unit vector,veccol!70!black]
\usetikzlibrary{angles,quotes} % for pic (angle labels)
\contourlength{1.3pt}

\usepackage{epsfig}
\usepackage{xcolor}
\usepackage{float,amsmath,amssymb}
\usepackage{graphicx}
\usepackage{url}
\usepackage{bbold}
\usepackage[utf8]{inputenc}
\usepackage{physics}
\usepackage[colorlinks=true,linkcolor=blue,citecolor=blue, urlcolor=blue]{hyperref}
\usepackage[]{mdframed}
\usepackage{bbold}
\newcommand{\pbpb}{$^{208}$Pb+$^{208}$Pb}
\newcommand{\xexe}{$^{129}$Xe+$^{129}$Xe}

\newcommand{\trento}{T$\mathrel{\protect\raisebox{-0.5ex}{R}}$ENTo}

\begin{document}

\title{Initial-state-driven spin correlations in high-energy nuclear collisions}

\author{Giuliano Giacalone}
\email{giuliano.giacalone@cern.ch}
\affiliation{Theoretical Physics Department, CERN, 1211 Geneva 23, Switzerland}

\author{Enrico Speranza}
\email{enrico.speranza@cern.ch}
\affiliation{Theoretical Physics Department, CERN, 1211 Geneva 23, Switzerland}

\begin{abstract}
In the study of spin-polarization phenomena in heavy-ion collisions, it is typically assumed that final-state particles are polarized through thermal vorticity and shear. In this sense, polarization is a final-state effect. Here, we propose a different mechanism. We postulate that the collision of spin-carrying nucleons generates an initial transverse spin density, inducing a net polarization of the QCD fireball along a random direction. If the net spin is conserved throughout the evolution of the fireball, the final-state particles should exhibit measurable polarization. Within a wounded nucleon picture, we estimate that initial-state fluctuations induce a net polarization of $\Lambda$ baryons which is around 1\% in central collisions and over 10\% in noncentral collisions, significantly exceeding the contributions from thermal vorticity and shear.  We introduce a two-particle angular correlation observable designed to reveal initial net-spin fluctuations, and emphasize the main signatures to look for in experiments. We argue that the discovery of these phenomena would have profound implications for nuclear structure and our understanding of spin in relativistic hydrodynamics.
\end{abstract}

\preprint{CERN-TH-2025-034}

\maketitle

\paragraph{\bf Introduction.}

There is a growing interest in the study of spin and related quantum phenomena in the context of high-energy physics and collider experiments. In particular, the observation of spin polarization of hadrons emitted in heavy-ion collisions has opened a new window onto the role of spin degrees of freedom in the dynamics of hot and dense QCD matter, or quark-gluon plasma (QGP) \cite{STAR:2017ckg,Adam:2018ivw,STAR:2019erd,ALICE:2019onw,ALICE:2019aid,STAR:2021beb,ALICE:2021pzu,STAR:2022fan,HADES:2022enx,CMS:2024byl,CMS:2025nqr}. This has triggered a vast research program aimed at elucidating the fundamental issue of the emergence of collective spin phenomena in relativistic quantum fluids \cite{Becattini:2020ngo,Speranza:2020ilk,Becattini:2024uha}.

The description of spin in heavy-ion collisions is typically based on the assumption of local equilibrium of the spin degrees of freedom, according to which final-state particles acquire polarization only in the presence of thermal vorticity or shear at the freeze-out of the QGP \cite{Liang:2004ph,Voloshin:2004ha,Betz:2007kg,Becattini:2007sr,Becattini:2013vja,Becattini:2013fla,Becattini:2015ska,Becattini:2016gvu,Karpenko:2016jyx,Pang:2016igs,Xie:2017upb,Becattini:2017gcx,Florkowski:2019qdp,Florkowski:2019voj,Zhang:2019xya,Becattini:2019ntv,Xia:2019fjf,Wu:2019eyi,Sun:2018bjl,Liu:2019krs}. In experiments, spin projections onto some fixed quantization axis are measured \cite{Niida:2024ntm}. Specifically, in the case of global and longitudinal polarization of $\Lambda$-hyperons, the chosen axes are the direction of the global angular momentum and the direction of the beam, respectively. Hydrodynamic simulations of the QGP including thermal vorticity and shear are capable of describing the qualitative features of the experimental results on $\Lambda$ polarization \cite{Ryu:2021lnx,Fu:2021pok,Becattini:2021suc,Becattini:2021iol,Palermo:2024tza}.
However, significant discrepancies remain, for instance in small systems \cite{CMS:2025nqr} or in the description of the spin alignment of vector mesons \cite{ALICE:2019aid,STAR:2022fan}, which motivates developments beyond the local equilibrium paradigm, such as relativistic spin hydrodynamics, see, e.g., \cite{Florkowski:2017ruc,Florkowski:2017dyn,Montenegro:2017rbu,Florkowski:2018fap,Hattori:2019lfp,Bhadury:2020puc,Weickgenannt:2020aaf,Shi:2020htn,Fukushima:2020ucl,Bhadury:2020cop,Gallegos:2020otk,Singh:2020rht,Bhadury:2021oat,Hu:2021pwh,Gallegos:2021bzp,Li:2020eon,She:2021lhe,Wang:2021ngp,Daher:2022xon,Cartwright:2021qpp,Hongo:2021ona,Weickgenannt:2022zxs,Weickgenannt:2022jes,Gallegos:2022jow,Cao:2022aku,Weickgenannt:2023bss,Becattini:2023ouz,Wagner:2024fhf,Wagner:2024fry,Singh:2024cub,Huang:2024ffg}. Most notably, the dynamics of spin in the \textit{initial conditions} of the collisions remains largely unexplored. 

In this Letter, we take a step in this direction. We go beyond the present paradigm where spin polarization in heavy-ion collisions emerges solely from the conversion of orbital angular momentum into local hydrodynamic fields. We argue that spin is present in the initial conditions of the collisions as it is carried by the nucleons within the colliding nuclei that participate in the interaction. To initialize spin in a realistic manner, we establish a model of an initial spin density that is affected by event-by-event fluctuations \cite{Alver:2010gr,Giacalone:2023hwk} associated with the finite number of interacting nucleons.

Moreover, we argue that the fluctuations of the initial net spin of the fireball induce polarization in the produced $\Lambda$ baryons and discuss an experimental protocol to measure this phenomenon. This goes beyond the aforementioned analyses measuring polarization with respect to a fixed direction, as we propose a frame-independent observable that measures the spin correlation of $\Lambda$ pairs, namely $\langle \cos (\Delta \theta) \rangle$, where $\Delta \theta$ is the relative angle between daughter protons, where the average is performed over all pairs in a heavy-ion collision (or \textit{event)}. It is interesting to note that, in the scenario in which the two decaying fermions form a two-qubit state, this observable provides a measurement of their quantum entanglement, as recently demonstrated by the analysis of top-antitop decays in proton-proton collisions at the Large Hadron Collider (LHC) \cite{ATLAS:2023fsd,CMS:2024pts,Afik:2020onf,Afik:2022kwm}. Here, we argue that the discovery of initial-spin fluctuations in heavy-ion collisions would similarly shed light on nontrivial spin-spin correlations in the wave functions of the colliding ions.

\begin{figure}[t]
    \centering
    \includegraphics[width=.935\linewidth]{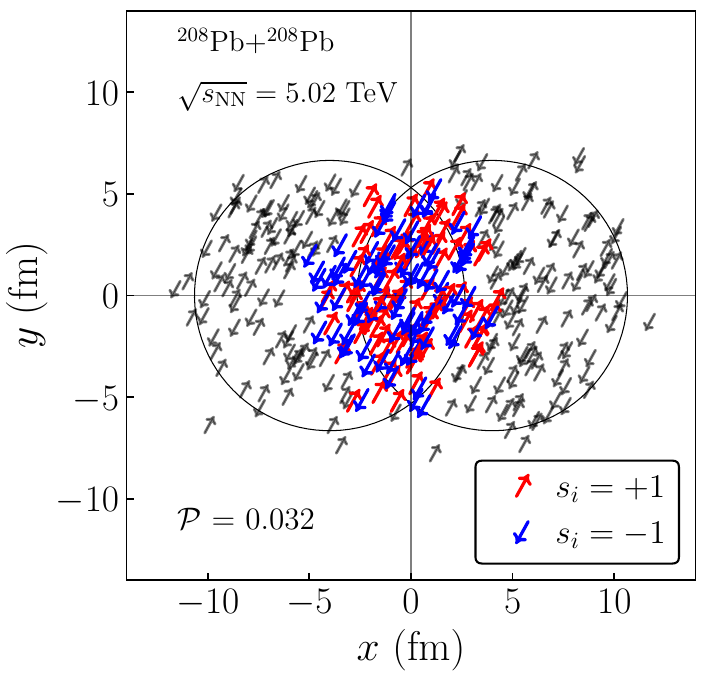}
    \caption{ Transverse view of a Pb+Pb collision at $\sqrt{s_{\rm NN}}=5.02$ TeV. Each nucleon within the colliding ions is associated with a random spin projection, either up or down. As total angular momentum must vanish for $^{208}$Pb, each ion is constituted of $A/2$ spin-up and $A/2$ spin-down nucleons. The spins are globally oriented in space, following the random quantization axis of the angular momenta of the colliding nuclei. The event shown corresponds to $\int_{\bf x} \mathcal{S}({\bf x})=6\,\hbar/2$ and $\mathcal{P}=0.032$.  }
    \label{fig:1}
\end{figure}

\paragraph{\bf Model of spin density and fluctuations.}

We introduce an initial spin density with realistic properties. Our main point is that such a quantity presents large event-by-event fluctuations. For the definition of the energy density of the fireball produced in a heavy-ion collision, it is crucial to consider the effect of event-by-event fluctuations in the spatial position of nucleons, which dominate the observed radial and anisotropic flow fluctuations \cite{Heinz:2013th,Bozek:2017elk,Giacalone:2023hwk,Schenke:2020uqq,Ollitrault:2023wjk}. Consequently, as nucleons carry spin, it is natural to expect that the randomness in the number of nucleons that participate in the collisions will also yield spin fluctuations in the initial conditions.

The way in which we introduce a fluctuating initial spin density is depicted in Fig.~\ref{fig:1}. In a Glauber-type description \cite{Miller:2007ri,Loizides:2014vua,dEnterria:2020dwq}, we consider the colliding ions as bunches of $A$ nucleons ($A=208$ for lead nuclei) whose positions are random in every event and independently sampled from a Woods-Saxon density profile. Each nucleon carries an intrinsic spin in agreement with the ground-state wave function of the ion which, within this framework, is understood to collapse at the time of collision. For $J=0$ nuclei like $^{208}$Pb, there must be $A/2$ nucleons with spin up and $A/2$ nucleons with spin down. We consider that the quantization axis of the spin has an orientation that is random with respect to the direction of the impact parameter, and that this direction is the same for both the incoming ions. For a collision at impact parameter $b=8$~fm, two nuclei prepared this way are shown in Fig. \ref{fig:1}. The collision occurs between pairs of nucleons, which defines the number of participants, $N_{\rm part}$. The participant nucleons are highlighted with colors (red for spin up, and blue for spin down). From $N_{\rm part}$ and the spins, we construct the spin density at mid-rapidity
    \begin{equation}
    \label{eq:Sdens}
        \mathcal S ({\bf x}) \equiv S_0 \frac{ \hbar } {2} \sum_{i=1}^{N_{\rm part}}s_i \, w_s({\bf x}- {\bf x}_i),
    \end{equation}
    where ${\bf x}$ is a coordinate in the transverse plane, $w_s({\bf x}-{\bf x}_i)$ is a two-dimensional profile which tells us how the spin is distributed around the participant location, $s_i$ is the participant's spin projection onto the chosen quantization axis, which takes values $s_i = \pm 1$, and $S_0$ is a (dimensionless) normalization. An example of $\mathcal{S}({\bf x})$ corresponding to the collision in Fig.~\ref{fig:1} is shown in Fig.~\ref{fig:S1} in the Supplemental Material. The resulting transverse density is reminiscent of recent evaluations of conserved-charge density distributions in \pbpb{} collisions \cite{Martinez:2019jbu,Carzon:2019qja,Carzon:2023zfp}.
\begin{figure}[t]
    \centering
    \includegraphics[width=.95\linewidth]{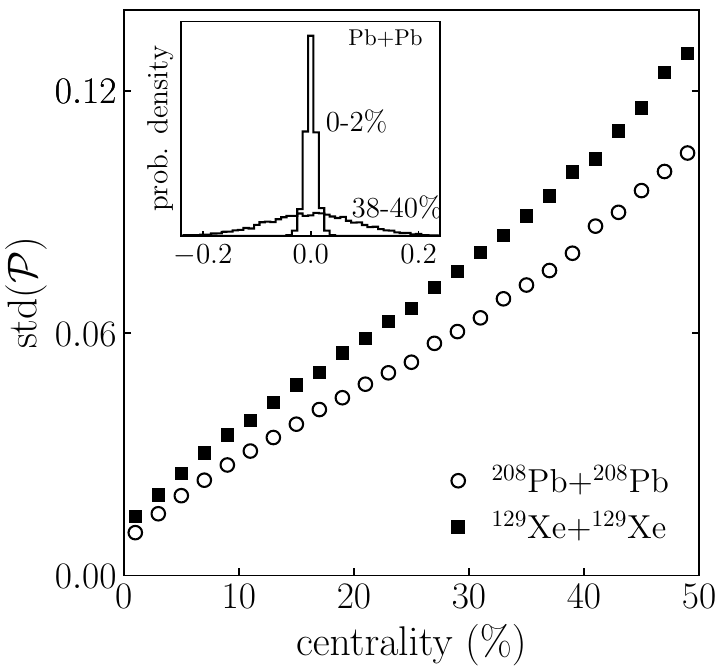}
    \caption{Standard deviation of the polarization parameter, $\mathcal{P}$, as defined by Eq.~(\ref{eq:stdP}), in \pbpb{} collisions (empty circles) and \xexe{} collisions (full squares) as a function of the collision centrality. Statistical uncertainties are smaller than the symbol sizes. The inset displays the probability density of $\mathcal{P}$ in two centrality classes for \pbpb{} events.}
    \label{fig:2}
\end{figure}

    With this spin profile, we assign to the event a certain degree of polarization, given by the net-spin of the initial condition. From the total spin at midrapidity
    \begin{equation}
    \label{eq:Pnet}
       \mathcal{S}  =  \int d^2{\bf x}\, \mathcal{S}({\bf x}) =  S_0  \frac{\hbar}{2} \sum_{i=1}^{N_{\rm part}} s_i,
    \end{equation}
    where we have considered $\int_{\bf x} w_s({\bf x}-{\bf x}_i)=1$, we define the polarization of the event
    \begin{equation}
        \mathcal{P} = \frac{1}{N_{\rm part}} \sum_{i=1}^{N_{\rm part}} s_i, \hspace{30pt} -1\leq \mathcal{P} \leq 1.
    \end{equation} 
 In this picture, an average over events leads to $\langle \mathcal{P} \rangle = 0$.  On the other hand, the standard deviation of the distribution is nontrivial as it follows the event-by-event fluctuations of $N_{\rm part}$. To show this, we perform minimum bias \pbpb{} collisions, and compute the value of $\mathcal{P}$ event-to-event. We evaluate its standard deviation, 
\begin{equation}
\label{eq:stdP}
    {\rm std}(\mathcal{P}) = \sqrt{\langle \mathcal P^2 \rangle},
\end{equation}
as a function of the collision centrality, defined from the total entropy of the collisions following the default \trento{} model \cite{Moreland:2014oya}. The result is shown in Fig.~\ref{fig:2}. We see, as expected, that the standard deviation of the polarization increases by one order of magnitude from central to peripheral collisions, in agreement with an $1/\sqrt{N_{\rm part}}$ scaling. We repeat the calculation for collisions of $^{129}$Xe nuclei. Owing to the smaller number of participants, the value of ${\rm std}(\mathcal P)$ in \xexe{} collisions is larger than in \pbpb{} collisions by about a factor $\sqrt{208/129}\approx1.27$ \cite{Giacalone:2017dud} across the centrality range.

With this picture in mind, our claim is that, in one event, the particles emitted from the QGP are polarized along some random vector, $\vec{\mathcal{P}}=\mathcal{P} \,\hat{n}$, where $\hat{n}$ is a unit vector which in Fig. \ref{fig:1} gives the direction of the chosen quantization axis. Consider the case of $\Lambda$ hyperons. If $\mathcal{P}=+1(-1)$, then all $\Lambda$ have polarization up(down) with respect to the direction of $\vec{\mathcal{P}}$. If $\mathcal{P}=0$, up and down spins are equally probable, and so on.  One comment is in order. Maximal values of the initial net spin are of the order of $100\hbar/2$. In principle, this is much smaller than the orbital angular momentum of the fireball, which is estimated to be $10^{4-5} \hbar$ \cite{Becattini:2015ska,Karpenko:2016jyx,STAR:2017ckg}. However, what matters for the polarization of the final-state $\Lambda$ hyperons is the spin which is induced by the conversion of orbital angular momentum into the vorticity, $\omega$, of the QGP. The relevant parameter for the polarization of the final-state hadrons is $\omega/T$, which in collisions at top Relativistic Heavy Ion Collider (RHIC) and LHC energies is of order $10^{-3}$ or smaller. On the other hand, as shown in Fig.~\ref{fig:2} the value of $\mathcal{P}$ can fluctuate up to several percents. Hence, initial-state spin fluctuations yield much higher polarization signals than the vorticity and the shear of the fireball \cite{Pang:2016igs,Palermo:2024tza}.

However, the observables so far analyzed to reveal polarization phenomena in heavy-ion collisions are computed with respect to some fixed axis that has no correlations with the direction of $\vec{\mathcal{P}}$. As a consequence,  the contribution of the initial-state polarization vanishes in the event averages \cite{footnote}. We introduce now an observable quantity to reveal the existence of an event-by-event polarization, $\mathcal{P}$, and therefore the potential signatures of initial-state spin fluctuations.

\paragraph{\bf Measuring spin-spin correlations.}   

We propose a new measure of the correlation of spins between different $\Lambda$-baryons.  Our rationale is the following.

\begin{figure}[t]
    \centering
    \includegraphics[width=.7\linewidth]{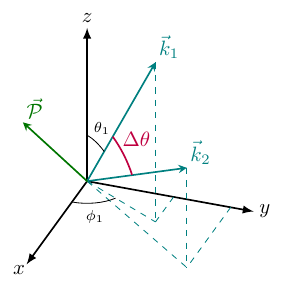}
    \caption{The decay protons in the rest frame of the parent hyperons are emitted with momenta $\vec k_1$ and $\vec k_2$. In the zero-momentum frame of the $\Lambda$ pair, the $z$ axis denotes the direction of their momenta. The relative angle is given by $\cos (\Delta \theta) = \hat{k}_1 \cdot \hat{k}_2$.  The direction of the polarization vector, $\vec{\mathcal{P}}$, cannot be determined in the experiment.}
    \label{fig:3}
\end{figure}

To build some intuition, we first show how one would determine that $\langle \mathcal {P} \rangle$ vanishes in an average over events. We depict in Fig.~\ref{fig:3} the decay of a $\Lambda$ hyperon.  We denote by $z$ the direction of the momentum of the $\Lambda$, with $\vec k$ the momentum of the daughter proton, and with $\vec{\mathcal{P}}$ the event-polarization vector.  Consider many such $\Lambda$ produced in one event. If one knew the direction of $\vec{\mathcal{P}}$, then measuring the angular distribution of the $\Lambda$ weak decay,
\begin{equation}
    \frac{dN}{d\cos\theta} \propto 1+ \alpha \mathcal{P} \,\hat{n} \cdot \hat{k}  = 1 + \alpha \mathcal{P} \cos\theta ,
\end{equation}
where $\alpha$ is the known weak-decay parameter \cite{ParticleDataGroup:2024cfk} and $\theta$ is the angle between the fixed quantization axis $\hat{n}$ and $\hat{k}=\vec{k}/|\vec{k}|$, would give access to the value of $\mathcal{P}$
\begin{equation}
    \mathcal{P} = \frac{3}{\alpha} \frac{ \int d\cos \theta \frac{dN}{d \cos \theta}\cos\theta } { \int d\cos\theta \frac{dN}{d \cos \theta} } \equiv \frac{3}{\alpha} \langle \cos \theta \rangle.
\end{equation}
This is the prescription followed in the analysis of the global $\Lambda$ polarization. However, in the picture of Fig.~\ref{fig:1} the sign of $\mathcal{P}$ is random in all events for unpolarized colliding ions, such that the event-average polarization vanishes:
\begin{equation}
   \langle \mathcal{P} \rangle  = \frac{3}{\alpha} \left \langle \frac{ \int d\cos \theta \frac{dN}{d \cos \theta}\cos\theta } { \int d\cos\theta \frac{dN}{d \cos \theta} } \right \rangle \equiv \frac{3}{\alpha} \langle  \! \langle \cos \theta \rangle \! \rangle = 0,
\end{equation}
Therefore, our goal is to measure the \textit{fluctuation} of $\mathcal{P}$.

Consider a pair of $\Lambda$ observed in the lab frame in a collision event. The goal is to define a common Cartesian coordinate for both decay protons. We proceed in analogy with the study of top-antitop spin correlation as explained in Sec. 2 of Refs. \cite{Baumgart:2012ay}. As a first step, we perform a boost from the lab frame to the zero-momentum frame of the $\Lambda$ pair, where their momenta are back-to-back. This enables us to uniquely identify a direction (denoted by $z$ in Fig. \ref{fig:3}) along which we subsequently boost to go to the rest frame of the $\Lambda$, where we measure the momenta of the decay protons, respectively $\vec{k}_1$ and $\vec{k}_2$. From the directions of these protons, we evaluate the angular average $\langle \cos (\Delta \theta) \rangle$ over all distinct pairs, with $\cos \left ( \Delta \theta \right ) = \hat{k}_1\cdot \hat{k}_2$. 
Finally, we average over events. This leads to a mean-squared polarization
\begin{equation}
    v_\Lambda^2  \equiv   \frac
{9}{\alpha_1 \alpha_2} \left \langle \! \left \langle  \cos (\Delta \theta)  \right \rangle \!  \right \rangle = \langle \mathcal{P}^2 \rangle ,   
\label{corrdelta}
\end{equation}
where we make a distinction between the decay constants to include the case of a correlation of $\Lambda$ and $\bar{\Lambda}$. The crucial property of this quantity is that it is frame-independent: it isolates the fluctuations of the magnitude of  $ \vec {\mathcal{P}}$ without requiring any knowledge of the \textit{direction} of this vector in the events, which cannot be determined experimentally (see the Supplemental Material). 

The result shown in Fig.~\ref{fig:2} gives, thus, the first prediction for the centrality and system-size dependence of the correlator \eqref{corrdelta}, in the limit where the spacetime evolution of the spin density does not alter the value of $\mathcal{P}$ (implying, as we discuss below, that the total spin is conserved from initial to final state). Therefore, our prediction should be tested against experimental data \cite{nonflow}. 

\paragraph{\bf Connection to relativistic spin hydrodynamics.}
This work presents the first prescription for defining an initial spin density in a realistic model of heavy-ion collisions. It is important to connect this advancement to the current understanding of relativistic spin hydrodynamics. 

The equations of motion of relativistic spin hydrodynamics in the case of an uncharged fluid are given by the conservation of energy, momentum, and total angular momentum. In our model, we assume that the total spin is conserved throughout the evolution of the system, such that the conservation of total angular momentum amounts to the conservation of spin \cite{Florkowski:2017ruc,Speranza:2020ilk} (see also the Supplemental Material). In the framework of quantum kinetic theory, this amounts to neglecting non-local collisions \cite{Weickgenannt:2020aaf,Weickgenannt:2022zxs} or to taking the standard relaxation time approximation for the collision term \cite{Bhadury:2020puc,Bhadury:2020cop}. Therefore, the equations of motion of relativistic spin hydrodynamics are given by:
\begin{equation}
    \partial_\mu T^{\mu\nu} =0, \qquad
    \partial_\lambda S^{\lambda,\mu\nu}=0,
    \label{cons2}
\end{equation}
where $T^{\mu\nu}$ is the energy-momentum tensor and $S^{\lambda,\mu\nu}$ is the so-called spin tensor, defined to be antisymmetric in the last two indices $S^{\lambda, \mu\nu}=-S^{\lambda, \nu\mu}$. We note that the definitions of $T^{\mu\nu}$ and $S^{\lambda, \mu\nu}$ are ambiguous, as one can always redefine the currents through so-called pseudo-gauge transformations that leave the total charges invariant \cite{Hehl:1976vr,Becattini:2018duy,Speranza:2020ilk,Buzzegoli:2024mra}. Therefore, caution should be taken when identifying $S^{\lambda,\mu\nu}$ as the current that carries information only on spin. Nonetheless, we can give an operational procedure that allows one to identify the components of $S^{\lambda, \mu\nu}$ with the spin density defined in Eq.~\eqref{eq:Sdens}.

In general, from the spin tensor one defines a relativistic spin density $S^{\mu\nu}=-S^{\nu\mu}=u_\lambda S^{ \lambda , \mu\nu}$, where $u^\lambda$ is the fluid velocity. Among the six independent components of $S^{\mu\nu}$, we make the assumption that, in the initial state, only the following three independent components \cite{maxwell}
\begin{equation}
\label{spinmu}
    S^{\mu}=\frac12 \epsilon ^{\mu\nu\alpha\beta}S_{\nu\alpha}u_\beta
\end{equation}  
are nonvanishing, where $S^\mu u_\mu=0$. It is thus natural to identify the spin density defined in Eq.~\eqref{eq:Sdens} as the components of Eq.~\eqref{spinmu} in the transverse plane, where it is assumed in the initial state the fluid velocity is given by $u^\mu=(1,0,0,0)$ at mid-rapidity. Therefore, we identify
\begin{equation}
    S^i \equiv n^i S ~ [{\rm \hbar/fm^{3}}],
\end{equation}
where $n^i$ is the unit vector previously used to define $\vec{\mathcal{P}}$.
Finally, we define an initial  density of spin at midrapidity,
\begin{equation}
 \lim_{\tau \rightarrow 0^+} \tau S   =    \mathcal{S}({\bf x}) ~[{\rm \hbar /fm^{2}}] ,
\end{equation}
where ${\bf x}$ is a coordinate in the transverse plane. Thus, this procedure provides a prescription for defining the initial conditions of relativistic spin hydrodynamic simulations, such as those carried out in, e.g., \cite{Singh:2024cub}.

\paragraph{\bf Spin correlations and nuclear structure.}
Collective flow measurements in heavy-ion collisions probe many-body properties, in particular, spatial correlations of nucleons in the wave functions of the colliding ions \cite{Giacalone:2023hwk}. We assess now what spin-correlation measurements probe in terms of nuclear structure. 

As explained in the Supplemental Material, measuring the imprints of initial-spin fluctuations probes the nuclear two-body density in spin space, formally defined via:
\begin{equation}
    \rho^{(2)}(s_1,s_2) = \sum_{s_3 \ldots s_A} \int_{{\bf x}_1, \ldots, {\bf x}_A} |\Psi({\bf x}_1, \ldots, {\bf x}_A, s_1, \ldots, s_A)|^2.
\end{equation}
If the number of participants in either nucleus is a  fixed value $N$, and considering central collisions where $N$ is close to the nuclear mass number, the variance of $\mathcal{P}$ reads:
\begin{equation}
    \langle \mathcal{P}^2 \rangle = N \sum_i \rho^{(1)}(s_i) s_i^2 + N(N-1) \sum_{ij} \rho^{(2)}(s_i,s_j)\,s_i\,s_j  .
\end{equation}
Here, as done in Fig.~\ref{fig:1}, we sample the spin of the nucleons independently, such that effectively $\rho^{(2)}(s_1,s_2) = \rho^{(1)}(s_1) \rho^{(1)}(s_2)$. Due to the symmetry of the one-body spin density, we are left with:
\begin{equation}
  \langle \mathcal{P}^2 \rangle =  N \sum_{s_i=\pm1} \rho^{(1)}(s_i) s_i^2,
\end{equation}
meaning that in the present model we do not implement any non-trivial two-body spin-spin correlation. However, nuclei are typically highly deformed objects, and we expect more sophisticated input from \textit{ab initio} nuclear structure calculations \cite{Hergert:2020bxy,Frosini:2021sxj,Tichai:2023epe,Gnech:2023prs,Arthuis:2024mnl,Giacalone:2024luz,Sun:2024iht,Lee:2025req} to present such effects. This potentially opens up a new way to study features in the structure of the colliding nuclei that, to the best of our knowledge, cannot be accessed by any other nuclear experiments at lower energies. 

\paragraph{\bf Summary and outlook.}

We have introduced a model of the initial spin density in high-energy nuclear collisions, and we have argued that it presents event-by-event fluctuations due to the finite $N_{\rm part}$. The initial net spin polarizes the fireball along some random direction, resulting in a polarization of the emitted $\Lambda$, which can be revealed via the two-particle correlation observable $v_\Lambda^2$.  These initial-state-driven spin effects should depend little on beam energy, and present a specific system-size scaling. In \pbpb{} collisions, following the variations of $N_{\rm part}$, the fluctuation of $\mathcal{P}$ varies from about 1\% in central to over 10\% in peripheral collisions. The observation of these patterns would shed light on unexplored initial conditions for relativistic spin hydrodynamics, as well as on two-body spin correlations in the colliding wave functions. Therefore, we urge the experimental collaborations to carry out the proposed measurement.

As an outlook, the discovery of initial-state spin fluctuations would pave the way for a number of studies. On the theory side, the identification of initial-state spin-related phenomena would allow us to thoroughly test the hypothesis of total spin conservation and, therefore, to benchmark theories of spin evolution and relaxation \cite{Kapusta:2019sad,Kapusta:2019ktm,Kapusta:2020npk,Hongo:2022izs,Hidaka:2023oze}. Experimentally, a new class of observables, accessible through high-statistics \pbpb{} runs from LHC Run 3 and Run 4, could be explored by correlating the spin of the $\Lambda$ with the momenta of the soft particles emitted in the collisions, as also suggested by a recent hydrodynamic study \cite{Alzhrani:2022dpi}. In particular, that would probe spin-position correlations in the ground states of the colliding nuclei. These searches can be refined in the future by the availability of a polarized gas for fixed-target experiments at the LHCb detector \cite{Aidala:2019pit}, which would allow us to test more directly whether the spin of the colliding nucleons affects the spin of the final-state hadrons. 

\paragraph{\bf Acknowledgments.} 
We thank Yoav Afik, Jo\~ao Barata, Juan R. M. de Nova,  Umut Gursoy, Tao Han, Alessandro Lovato,  Govert Nijs, Jorge Noronha, Jean-Yves Ollitrault, Enrique Rico, Wilke van der Schee, Ilya Selyuzhenkov, Urs Wiedemann, and Raul Wolters for useful discussions. E.S. has received funding from the European Union’s Horizon Europe research and innovation program under the Marie Sk\l odowska-Curie grant agreement No. 101109747.

\section{Supplemental Material}

\subsection{Calculation of $v_\Lambda^2$}

The four-fold angular distribution of two protons emitted by the weak decay of a pair of $\Lambda$ is given by
\begin{equation}
\label{fourang}
    \frac{d^4N}{d\Omega_1 d\Omega_2} \propto 1+ \alpha_1 \vec{\mathcal{P}}_1 \cdot \hat{k}_1 + \alpha_2 \vec{\mathcal{P}}_2 \cdot \hat{k}_2 + \sum_{i,j} C^{ij} \hat{k}_1^i\hat{k}_2^j,
\end{equation}
where $d\Omega_{1,2}=d\cos\theta_{1,2}d\phi_{1,2}$ are the solid angles of emission of the protons in their parent $\Lambda$ rest frames, and $\hat{k}_{1,2}$ are the corresponding unit vectors of their momenta. The vectors $\vec{\mathcal{P}}_1$ and $\vec{\mathcal{P}}_2$ are the polarization vectors of the $\Lambda$ hyperons, while the components of the matrix $C^{ij}$ encodes the spin correlations. Note that the expression in Eq. \eqref{fourang} is fully general and was also used to study spin correlations in top-antitop states created in proton-proton collisions \cite{Bernreuther:2004jv,Baumgart:2012ay,Bernreuther:2015yna}. The quantities $\mathcal{P}^i_{1,2}$ and $C^{ij}$ depend on the specific choice of the coordinate systems. Following Ref. \cite{Baumgart:2012ay}, in order to define a frame common to both decay protons, we first boost to the zero-momentum frame of the $\Lambda$ pair and then boost along their momentum direction. For example, denoting such a direction by $z$, integrating Eq.~\eqref{fourang} over the azimuthal angles yields
\begin{align}
\nonumber     \frac{d^2N}{d\cos\theta_1\cos\theta_2} \propto & \,\, 1 + \alpha_1\mathcal{P}^z_1\cos \theta_1 + \alpha_2\mathcal{P}^z_2 \cos \theta_2 \\
    &+ C^{zz}\cos \theta_1\cos \theta_2,
\end{align}
from which one can extract the values of $\mathcal{P}^z_{1,2}$ and $C^{zz}$ with an appropriate angle average. 

However, it is possible to access frame-invariant quantities (i.e., that do not depend on the specific choice of the Cartesian axes in the $\Lambda$ pair rest frames) by recasting Eq. \eqref{fourang} in the form
\begin{equation}
\label{DD}
    \frac{dN}{d\cos(\Delta\theta)} \propto  1 + D \cos (\Delta\theta),
\end{equation}
where $D=(1/3)\text{Tr}(C^{ij})$ and $\cos (\Delta \theta) = \hat{k}_1 \cdot \hat{k}_2$.
The spin correlation coefficient can be obtained as
\begin{equation}
    D=3  \langle \cos (\Delta \theta) \rangle. 
\end{equation}
where the average over angles reads
\begin{equation}
    \langle \cos (\Delta \theta) \rangle= \frac{\int d\Omega_1 d\Omega_2 \frac{d^4N}{d\Omega_1 \Omega_2} \cos (\Delta \theta) }{\int d\Omega_1 d\Omega_2 \frac{d^4N}{d\Omega_1 \Omega_2}}. 
\end{equation}
We emphasize that this is a frame-invariant quantity, and that it was used in Refs.~\cite{Afik:2020onf,Afik:2022kwm} as a measure of the quantum entanglement of the top-antitop pair.

Let us now specialize this general discussion to the model introduced in the main text, where we have $\vec{\mathcal{P}}_1=\vec{\mathcal{P}}_2=\vec{\mathcal{P}}=\mathcal{P} \hat{n}$, which vanishes after the average over events. Furthermore, it is assumed that the decays of different $\Lambda$ are independent processes and, hence, the angular distribution \eqref{fourang} factorizes 
\begin{equation}
     \frac{d^4N}{d\Omega_1 d\Omega_2}=\frac{d^2N}{d\Omega_1} 
     \frac{d^2N}{d\Omega_2},
\end{equation}
with 
\begin{equation}
    \frac{d^2N}{d\Omega_{1,2}} \propto 1+ \alpha_{1,2} \vec{\mathcal{P}} \cdot \hat{k}_{1,2},
\end{equation}
which implies $C^{ij}=\alpha_1 \alpha_2 \mathcal{P}^i \mathcal{P}^j$, such that
\begin{equation}
    D=\frac{\alpha_1 \alpha_2}3 \mathcal{P}^2.
\end{equation}
Hence, the quantity we want to measure is given by the event average of the spin correlation coefficient, namely,
\begin{align}
    v_\Lambda^2 \equiv  \frac{9}{\alpha_1 \alpha_2} \langle\!\langle \cos (\Delta \theta) \rangle\!\rangle = \frac{3}{\alpha_1 \alpha_2} \langle D \rangle = \langle \mathcal{P}^2 \rangle  .
\end{align}
The conclusion is that this observable isolates the event-by-event fluctuation of the \textit{magnitude} of $\vec{\mathcal{P}}$  without requiring the knowledge of its \textit{direction} in every event.

\subsection{Basics of relativistic spin hydrodynamics}
The equations of motion of relativistic spin hydrodynamics are the conservation of energy, momentum, and total angular momentum respectively given by 
\begin{equation}
    \partial_\mu T^{\mu\nu} =0, \qquad
    \partial_\lambda J^{\lambda,\mu\nu}=0,
    \label{cons222}
\end{equation}
where $T^{\mu\nu}$ is the energy-momentum tensor and $J^{\lambda,\mu\nu}$ the total angular-momentum tensor. The latter can, in turn, be decomposed into 
\begin{equation}
\label{decomp222}
J^{\lambda,\mu\nu}=L^{\lambda,\mu\nu}+S^{\lambda, \mu\nu},
\end{equation}
with $L^{\lambda,\mu\nu}=x^\mu T^{\lambda\nu}-x^\nu T^{\lambda\mu}$ and $S^{\lambda,\mu\nu}$. Inserting Eq.~\eqref{decomp222} into the second equation in \eqref{cons222}, and using the first equation in \eqref{cons222}, we obtain 
\begin{equation}
\label{scons222}
    \partial_\lambda S^{\lambda, \mu\nu}=T^{\nu\mu}-T^{\mu\nu},
\end{equation}
which shows that the spin tensor is not conserved if there is an antisymmetric part of the energy-momentum tensor. The system 
\eqref{cons222} consists of ten independent coupled partial differential equations, with explicit forms determined by the specific formulation considered. In the simplest case of ideal spin hydrodynamics \cite{Florkowski:2017ruc}, the independent fields one solves for are the energy density $\varepsilon$, the three independent components of the fluid velocity $u^\mu$ $(u^\mu u_\mu=1)$, together with the six independent components of the spin potential $\Omega^{\mu\nu}=-\Omega^{\nu\mu}$. The latter fields are additional degrees of freedom in spin hydrodynamics that correspond to the Lagrange multipliers associated with the total angular momentum conservation. In general, $S^{\mu\nu}=S^{\lambda,\mu\nu}u_\lambda$  is a function of $\Omega^{\mu\nu}$, where the explicit form is given by an equation of state. Thus, one can eventually relate the spin density in \eqref{spinmu} with $\Omega^{\mu\nu}$.

\subsection{Spin-spin correlations in the colliding nuclei}

\begin{figure}[t]
    \centering
    \includegraphics[width=0.9\linewidth]{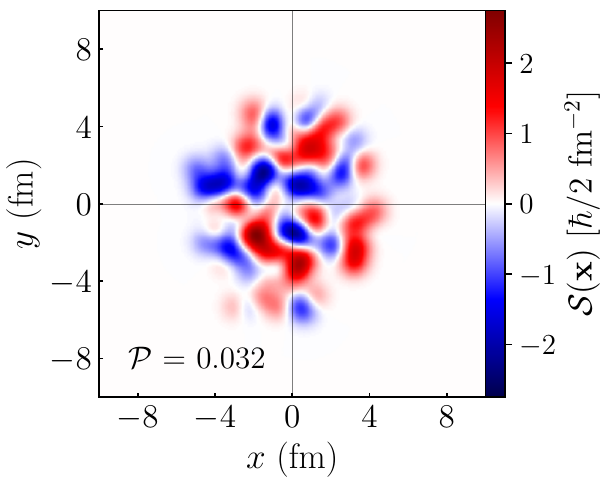}
    \caption{The spin density, $\mathcal{S}({\bf x})$, as defined via Eq.~(\ref{eq:SdensSM}), corresponding to the same collision event displayed in Fig.~\ref{fig:1}. For this plot, we consider that the spin profile of an individual participant, $w_s({\bf x})$, is a two-dimensional Gaussian of width 0.5 fm, and that $S_0=1$.}
    \label{fig:S1}
\end{figure}

We recall the definition of the initial spin density:
\begin{equation}
\label{eq:SdensSM}
    \mathcal{S}({\bf x})= S_0 \frac{\hbar}{2} \sum_{i=1}^{N_{\rm part}} s_i w_s({\bf x} - {\bf x}_i).
\end{equation}
With $s_i=\pm 1$, and $\int_{\bf x}  w_s ({\bf x})= 1$. One example of such density is shown in Fig.~\ref{fig:S1}. The dimensionless initial-state polarization parameter reads:
\begin{equation}
 \mathcal{P} = \frac{1}{N_{\rm part}} \sum_{i=1}^{N_{\rm part}} s_i.
\end{equation}
In this section, we study the statistical properties of the random number $\mathcal{P}$.

\subsubsection{The mean value}
In the event-average of the initial-state polarization:
\begin{equation}
    \langle \mathcal{P} \rangle = \biggl \langle  \frac{1}{N_{\rm part}} \sum_i^{N_{\rm part}} s_i \biggr \rangle ,
\end{equation}
both $N_{\rm part}$ and $s_i$ fluctuate. As done in Fig.~\ref{fig:2}, we focus on collisions at fixed total entropy, or fixed centrality, which amounts to a great extent to fixing the value of $N_{\rm part}$. In addition, we decompose the total number of participants as $N_{\rm part} = N_A + N_B$, where $N_A$ and $N_B$ label, respectively, the number of participants coming from the two colliding ions. This leads to:
\begin{equation}
\label{eq:avePAB}
    \langle \mathcal{P} \rangle =   \frac{1}{N_{\rm part}} \biggl \langle \sum_{i=1}^{N_{\rm A}} s_{i} + \sum_{i=1}^{N_{\rm B}} s_{i} \biggr \rangle.
\end{equation}
Now, let us for simplicity assume that $N_A$ is fixed in the class of considered events. The contribution to the average of $\mathcal{P}$ becomes:
\begin{equation}
    \biggl \langle \sum_i^{N_{\rm A}} s_{i} \biggr \rangle = \sum_{i}^{N_A} \langle s_i \rangle,
\end{equation}
where
\begin{equation}
    \langle s_i \rangle =  \sum_{s_i=\pm1} \rho_{\rm part}^{(1)}(s_i) \, s_i\, .
\end{equation}
Here $\rho^{(1)}_{\rm part}(s_i)$ is the one-body density of participant spins. If we further consider that the collisions are central enough such that $N_{\rm part}\approx 208$, we can replace  $\rho_{\rm part}^{(1)}(s_i)$ with the full particle density of the nucleus:
\begin{equation}
    \langle s_i \rangle \approx  \sum_{s_i=\pm1} \rho^{(1)}(s_i) \, s_i\, ,
\end{equation}
where we introduce the one-body density of the nuclear ground state in spin space:
\begin{equation}
    \rho^{(1)}(s_1) = \sum_{s_{2}, \ldots  ,s_A} \int_{{\bf r}_1, \ldots, {\bf r}_A} |\Psi({\bf r}_1, \ldots, {\bf r}_A, s_1, \ldots, s_A)|^2,
\end{equation}
where {\bf r} is a coordinate in three-dimensional space and with normalization
\begin{equation}
    \sum_{s_1=\pm1} \rho^{(1)}(s_1) = 1 \, .
\end{equation}
With this defined, we can see that the contributions to the average value of $\mathcal{P}$ in Eq.~(\ref{eq:avePAB}) vanish as a consequence of the symmetry of the one-body density:
\begin{equation}
  \langle s_i \rangle =  \sum_{s_i=\pm 1} \rho^{(1)}(s_i)\,s_i = 0,
\end{equation}
which leads to the expected result:
\begin{equation}
    \langle \mathcal P \rangle = 0.
\end{equation}
Note that this result does not change if we replace $\rho^{(1)}(s_i)$ with the original $\rho_{\rm part}^{(1)}(s_i)$, which must present the same reflection symmetry in the absence of any polarization of the colliding nucleons.

\subsubsection{The fluctuation}

We evaluate now the variance of the polarization parameter:
\begin{equation}
    {\rm var}(\mathcal{P}) = \langle \mathcal{P}^2 \rangle.
\end{equation}
Following the previous discussion, we can write:
\begin{align}
    \mathcal{P} &= \frac{1}{N_{\rm part}} \biggl ( \sum_{i=1}^{N_A} s_{i,A} + \sum_{i=1}^{N_B} s_{iB} \biggr ), \\
\nonumber    \mathcal{P}^2  &= \frac{1}{N_{\rm part}^2} \biggl ( 
\nonumber \sum_{i=1}^{N_A}\sum_{j=1}^{N_A} s_{i,A}s_{j,A} \\ 
& \hspace{50pt} + 2 \sum_{i=1}^{N_A}\sum_{j=1}^{N_B} s_{i,A} s_{j,B} 
  + \sum_{i=1}^{N_B}\sum_{j=1}^{N_B} s_{i,B} s_{j,B}  \biggr ).
\end{align}
Now we calculate $\langle \mathcal{P}^2 \rangle$.  Since $s_{i,A}$ and $s_{i,B}$ come from different nuclei, they are independent variables. Therefore:
\begin{equation}
  \langle  s_{i,A} s_{j,B} \rangle =   \langle  s_{i,A} \rangle \langle s_{j,B} \rangle = 0,
\end{equation}
and we are left with:
\begin{equation}
    \langle \mathcal{P}^2 \rangle = \frac{1}{N_{\rm part}^2} \biggl (  \sum_{i=1}^{N_A}\sum_{j=1}^{N_A} \langle s_{i,A}s_{j,A} \rangle +  \sum_{i=1}^{N_B}\sum_{j=1}^{N_B} \langle s_{i,B}s_{j,B} \rangle \biggr ).
\end{equation}
As done for the mean value of $\mathcal{P}$, we focus on one of the two correlators, which involves two participant nucleons coming from the same nucleus. Henceforth, we drop the $A$ subscripts and set $N_A = N$. In addition, we consider again very central collisions for which we can approximate the many-body participant densities, $\rho_{\rm part}^{(n)}(s_1\ldots s_n)$, with those of the actual nuclear ground state.
By doing so, there are $N$ terms with $i=j$ and $N(N-1)$ off-diagonal terms:
\begin{align}
\label{eq:sisj}
    \langle s_{i}s_{j} \rangle =  & \,\, N \sum_{s_i=\pm1} \rho^{(1)}(s_i) s_i^2  \nonumber\\
    &+ N(N-1) \sum_{\substack{s_i,s_j=\pm1 \\ i\neq j}} \rho^{(2)}(s_i,s_j)\,s_i\,s_j  .
\end{align}
where we introduce a two-body density in spin space:
\begin{equation}
    \rho^{(2)}(s_1,s_2) = \sum_{s_3\ldots s_A} \int_{{\bf r}_1, \ldots, {\bf r}_A} |\Psi({\bf r}_1, \ldots, {\bf r}_A, s_1, \ldots, s_A)|^2.
\end{equation}
This formula demonstrates that the fluctuation of $\mathcal{P}$ in the picture considered in this paper gives access to two-spin correlations in the nuclear ground states.

In the model considered in this article, we treat the spins of the nucleons as independent random variables. Therefore, we are effectively factorizing the density:
\begin{equation}
    \rho^{(2)}(s_1,s_2) = \rho^{(1)}(s_1) \rho^{(1)}(s_2).
\end{equation}
When this is plugged in Eq.~(\ref{eq:sisj}), one ends up with:
\begin{align}
 \nonumber   \langle s_{i}s_{j} \rangle = &  N \sum_i \rho^{(1)}(s_i) s_i^2  \\ 
\nonumber    & \hspace{20pt} + N(N-1) \sum_{i} \rho^{(1)}(s_i)\,s_i \sum_j \rho^{(1)}(s_j)\,s_j \\
    & =  N \sum_i \rho^{(1)}(s_i) s_i^2  ,
\end{align}
where we have used  $\sum_{s_i = \pm 1/2} \rho^{(1)}(s_i) s_i  = 0$.
Therefore, we conclude that $\langle s_{i} s_{j} \rangle$ is non-zero, such that
\begin{equation}
   \langle \mathcal{P}^2 \rangle  > 0,
\end{equation}
driven by the variance of the one-body density of the nucleus in spin space.

\bibliographystyle{apsrev4-1}
\bibliography{biblio}

\begin{thebibliography}{121}
\bibitem [{max()}]{maxwell}%
  \BibitemOpen
  \href@noop {} {}\bibinfo {note} {This is akin to decomposing the Maxwell tensor into an electric and a magnetic part, and by considering that the electric part vanishes, which is the approximation used in non-resistive magnetohydrodynamics \cite{Rezzolla_Zanotti_book}.}\BibitemShut {Stop}%
\end{thebibliography}

\end{document}